\documentclass{elsart}

\usepackage{graphicx}

\begin{document}

\begin{frontmatter}

\title{Twist of fractional oscillations}

\author{Aleksander~A.~Stanislavsky\corauthref{cor1}}
\corauth[cor1]{E-mail: alexstan@ri.kharkov.ua}

\address{Institute of Radio Astronomy, 4 Chervonopraporna
St., Kharkov 61002, Ukraine}

\begin{abstract}
Using the method of the Laplace transform, we consider fractional
oscillations. They are obtained by the time-clock randomization of
ordinary harmonic vibrations. In contrast to sine and cosine, the
functions describing the fractional oscillations exhibit a finite
number of damped oscillations with an algebraic decay. Their
fractional differential equation is derived.
\end{abstract}


\end{frontmatter}

\section{Introduction}
\thispagestyle{empty} According to fractional calculus the
harmonic oscillator is a particular case of the fractional
oscillator \cite{1,1a,1b}. It suffices to say that the fractional
oscillator equation is a generalization of the classical harmonic
equation by replacing the second-order derivative by a fractional
one. If the harmonic oscillator is described by elementary
harmonic functions, the fractional oscillator solution is
expressed in terms of Mittag-Leffler functions. The interrelation
between these oscillators has clarified with the consideration of
temporal subordination. Then the development of an oscillatory
process is governed by its own internal clock (operational time)
that is not synchronized with physical (deterministic) clock
\cite{1c}. As has been shown in \cite{2} that the fractional
oscillator results from an ensemble average of harmonic
oscillations under stochastic time arrow.  In fact, the fractional
oscillator can be considered as a time-clock randomization
(subordination) of the conventional harmonic oscillator. The new
time clock is the continuous limit of the discrete counting
process, when the time variable is a sum of random temporal
intervals belonging to an $\alpha$-stable distribution.  Hence
there exists a direct correspondence between the functions of both
oscillators.

In particular, the cosine function $\cos(t)$ is connected with the
one-parameter Mittag-Leffler function $E_\alpha(-t^\alpha)$. Its
features has been studied in detail (see, for example, \cite{1}
and references therein). Passing to the limit $\alpha\to 2$, this
Mittag-Leffler function transforms to $\cos(t)$. One of very
interesting properties of the one-parameter Mittag-Leffler
function is a finite number of zeros. Therefore, the function has
a finite number of damped oscillations with an algebraic decay.
For completeness it is necessary to investigate also the ``sine''
correspondence of fractional oscillations. The function is briefly
mentioned in~\cite{1b} without any detailed analysis. In this
connection it should be pointed out that the displacement of
harmonic oscillator and its momentum are written by means of
ordinary harmonic functions. Similar values (displacement and
momentum) may be formulated for fractional oscillator too. Their
consideration is of great interest. They are just expressed in
terms of Mittag-Leffler functions mentioned above. The aim of this
paper is to study their features in the context of fractional
oscillator.

The paper is organized as follows. In Section~\ref{par2} we start
our analysis with the time-clock randomization of the ordinary
harmonic oscillator. This allows us to use the integral relation
between the harmonic functions and the Mittag-Leffler functions of
the fractional oscillator. Our main interest will be focused on
the Mittag-Leffler function having the sine function as a limit.
Next we investigate its properties in Section~\ref{par3} and
derive a fractional differential equation for which this function
is its solution (Section~\ref{par4}). As a comparison, the
``cosine'' correspondence of fractional oscillations will be given
simultaneously. Section~\ref{par5} is devoted to a consideration
of fractional oscillation zeros.

\section{Probabilistic point of view}\label{par2}

If one randomizes the time clock in accordance with \cite{2}, the
fractional oscillations are written by means of
\begin{eqnarray}
A(t)&=&\int_0^\infty p^{S}(t,\tau)\,\cos\omega\tau\,d\tau=
E_{\alpha,1}(-\omega^2t^{\alpha})=E_{\alpha}(-\omega^2t^{\alpha})\,,
\nonumber\\ B(t)&=&\int_0^\infty
p^{S}(t,\tau)\,\sin\omega\tau\,d\tau=\omega t^{\alpha/2}
E_{\alpha,\,1+\alpha/2}(-\omega^2t^{\alpha})\,,\nonumber
\end{eqnarray}
where $0<\alpha<2$,
\begin{displaymath}
E_{\mu,\,\nu}(z)=\sum_{k=0}^{\infty}\frac{z^k}{\Gamma(\mu k+\nu)},
\qquad \mu,\nu>0,
\end{displaymath}
is the Mittag-Leffler function, $\Gamma(z)$ the gamma function,
and
\begin{displaymath}
p^{S}(t,\tau)=\frac{1}{2\pi j}\int_{Br} e^{st-\tau
s^{\alpha/2}}\,s^{-1+ \alpha/2}\,ds
\end{displaymath}
determines the probability to be at the internal time $\tau$ on
the real time $t$. Here ${\it Br}$ denotes the Bromwich path (a
line ${\it Re}\, s=\sigma$ with a value $\sigma\geq 1$), and
$j=\sqrt{-1}$.

Since the function $A(t)$ has been reviewed in \cite{1}, we devote
our attention more to the function $B(t)$. The latter has a
Laplace inversion integral
\begin{equation}
B(t)=\frac{\omega}{2\pi j}\int_{Br}
e^{st}\frac{s^{-1+\alpha/2}}{s^\alpha+\omega^2}\,ds\,.\label{eq1}
\end{equation}
The fractional oscillations exist for $1<\alpha<2$. It is that the
case is of interest to us and will be discussed below.

\begin{figure}
\centering
\includegraphics[width=12 cm]{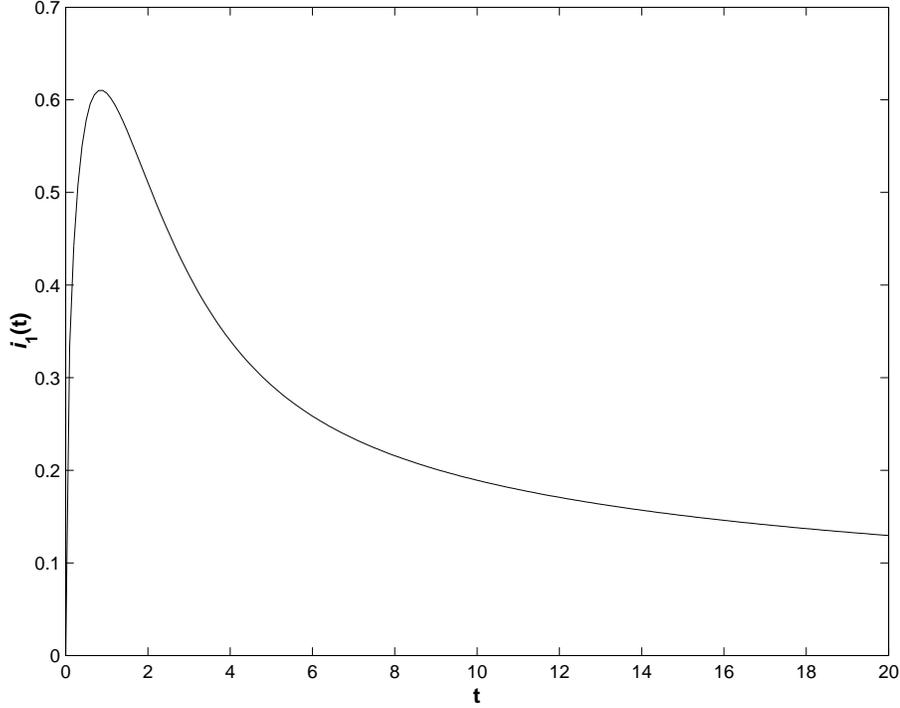}
\caption{\label{fig1}Fractional one-half-order integral of the
exponential function, $i_1(t)=J^{1/2} \exp(t)$\,.}
\end{figure}

For convenience one puts $\omega=1$, then one denotes $B(t)$ as
$i_\alpha(t)$. It should be pointed out that for $\omega=1$ the
function $A(t)$ is equal to $e_\alpha(t)$ according to the
notations of \cite{1}. As the Laplace image of the function
$e_\alpha(t)=\sum_{k=0}^\infty(-t^\alpha)^k/\Gamma(\alpha k+1)$ is
$s^{\alpha-1}/(1+s^\alpha)$, it is easy to determine the relation
between $e_\alpha(t)$ and $i_\alpha(t)$ via the fractional
integral of $\alpha/2$ order, namely
\begin{displaymath}
i_\alpha(t)=J^{\alpha/2}e_\alpha(t)=\frac{1}{\Gamma(\alpha/2)}\int_0^t
(t-\tau)^{-1+\alpha/2}\,e_\alpha(\tau)\,d\tau\,.
\end{displaymath}
Recall that for $\alpha=1,2$ the function $e_\alpha(t)$ takes the
form
\begin{displaymath}
e_1(t)=e^{-\,t},\qquad e_2(t)=\cos t\,.
\end{displaymath}
Define the same in reference to the function $i_\alpha(t)$ for
$\alpha=1,2$. At once it is obvious that $i_2(t)=\sin t$. Using
the Efros' theorem, we find $i_1(t)$ via the Laplace image
\begin{displaymath}
i_1(t)=\frac{1}{2\pi j}\int_{Br}
e^{st}\frac{ds}{\sqrt{s}\,(s+1)}=\frac{1}{\sqrt{\pi
t}}\int_0^\infty e^{-\tau^2/(4t)}\,\sin\tau\,d\tau=e^{-\,t}
\mathrm{erfi}(\sqrt{t})\,.
\end{displaymath}
Note that
$\mathrm{erfi}(x)=\frac{2}{\sqrt{\pi}}\int_0^x\exp(y^2)\,dy$ is
the imaginary error function \cite{2a}. It is a real-valued,
entire function defined by $\mathrm{erfi}(x)=\mathrm{erf}(jx)/j$.
In contrast to the ordinary error function
$\mathrm{erf}(x)=\frac{2}{\sqrt{\pi}}\int_0^x\exp(-y^2)\,dy$, the
imaginary error function is not bounded, but the function $i_1(t)$
is bounded. It tends to zero in the limit $t\to\infty$. Moreover,
the latter is very close to the Dawson's integral
$D(x)=e^{-\,x^2}\int_0^x\exp(y^2)\,dy$ \cite{2b}. In this
connection it should be mentioned the asymptotic behavior of the
imaginary error function
\begin{displaymath}
\mathrm{erfi}(x)=\frac{1}{\sqrt{\pi}}\,e^{\,x^2}\left(x^{-1} +
\frac{1}{2}\,x^{-\,3} + \frac{3}{4}\,x^{-\,5} +
\frac{15}{8}\,x^{-7} + \dots\right)\,.
\end{displaymath}
It may be advantageous for the estimation of $i_1(t)$ with $t\gg
1$. To sum up above is the numerical simulation of $i_1(t)$
represented in Fig.~\ref{fig1}.

\section{Decomposition}\label{par3}

The next important step consists in the decomposition
$i_\alpha(t)$ into two contributions. For this purpose we bow the
Bromwich path of integration into the equivalent Hankel path. Then
the loop will start from minus infinity along the lower side of
negative real axis, encircle $|s|=1$ counter-clockwise and end at
minus infinity along the upper side of the negative real axis. The
first contribution arises from two borders of the cut negative
real axis. Taking $s=re^{j\pi}$ along the upper border and
$s=re^{-j\pi}$ along the lower border, we get
\begin{displaymath}
h_\alpha(t)=\int_0^\infty e^{-\,rt}\,V_\alpha(r)\,dr
\end{displaymath}
with
\begin{displaymath}
V_\alpha(r)=\frac{1}{\pi}\,\frac{r^{-1+\alpha/2}(1-r^\alpha)
\sin(\pi\alpha/2)}{r^{2\alpha}+2r^\alpha\cos(\pi\alpha)+1}\,.
\end{displaymath}
The part $h_\alpha(t)$ is not completely monotonic, as the
function $V_\alpha(r)$ takes on both positive and negative values
(see Theorem~7 in \cite{2c}). It becomes vanishingly small with
$t$ tending to infinity. The second contribution is calculated by
means of residues. The poles $s_0=\exp(j\pi/\alpha)$ and
$s_1=\exp(-j\pi/\alpha)$ give
\begin{displaymath}
q_\alpha(t)=\frac{2}{\alpha}\,e^{\,t\cos(\pi/\alpha)}\sin
\left[t\sin\Bigl(\frac{\pi}{\alpha}\Bigr)\right].
\end{displaymath}
This part clearly demonstrates an oscillatory character. Due to
it, the oscillatory behavior is carried over to the function
$i_\alpha(t)$ itself.

\begin{figure}
\centering
\includegraphics[width=12 cm]{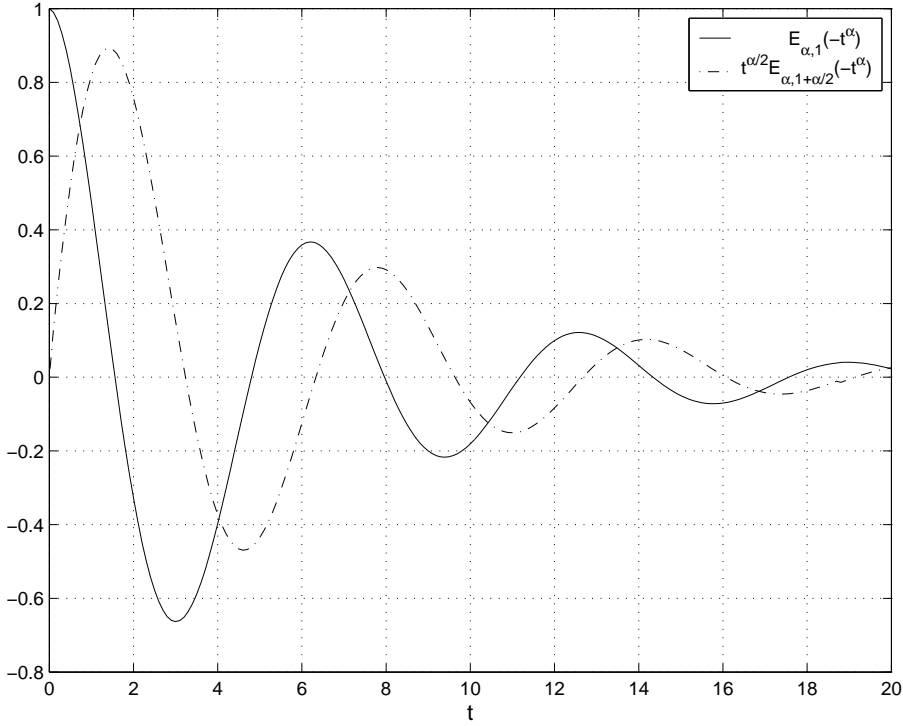}
\caption{\label{fig2}Fractional oscillations for $\alpha=1.8$\,.}
\end{figure}

It is useful to remark that the functions $i_\alpha(t)$ resembles
$e_\alpha(t)$. So for $1<\alpha<2$ the function $e_\alpha(t)$ also
consists of two parts, $e_\alpha(t)=f_\alpha(t)+g_\alpha(t)$. From
\cite{1} it follows that the monotonic part is
\begin{displaymath}
f_\alpha(t)=\int_0^\infty e^{-\,rt}\,K_\alpha(r)\,dr\,,
\end{displaymath}
where
\begin{displaymath}
K_\alpha(r)=\frac{1}{\pi}\,\frac{r^{\alpha-1}
\sin(\pi\alpha)}{r^{2\alpha}+2r^\alpha\cos(\pi\alpha)+1}\,.
\end{displaymath}
According to \cite{2c}, the function $f_\alpha(t)$ is complete
monotonic. The second contribution of $e_\alpha(t)$ takes the
oscillatory form
\begin{displaymath}
g_\alpha(t)=\frac{2}{\alpha}\,e^{\,t\cos(\pi/\alpha)}\cos
\left[t\sin\Bigl(\frac{\pi}{\alpha}\Bigr)\right].
\end{displaymath}
To have a clear idea of $i_\alpha(t)$ and $e_\alpha(t)$, their
pictures are represented in Fig.~\ref{fig2}.

\section{Equations of fractional oscillations}\label{par4}

The first approach to fractional oscillations had enough a formal
character. It proceeded from the simple change of the second
derivative in the harmonic oscillator equation to the derivative
of a fractional one. In other words, such a equation was
postulated. After solving it the fractional oscillation features
were established. Nevertheless, there is an alternative way to
fractional calculus \cite{3}. The probabilistic point of view
shows that the derivative of fractional order is connected with
$\alpha$-stable probability distributions \cite{4}. Starting from
the analysis of stochastic random processes, this permits ones to
derive the fractional differential equations responsible for
fractional oscillations. Here we intend to give new examples
demonstrating this approach.

Let the sign $\div$ be for the juxtaposition of a function
depending on $t$ with its Laplace image depending on $s$, namely
\begin{displaymath}
u(t)\quad\div\quad \bar u(s)=\int_0^\infty e^{-st}\,u(t)\,dt.
\end{displaymath}
The reader is reminded that
\begin{displaymath}
e_\alpha(t)\quad\div\quad\frac{s^{\alpha-1}}{s^\alpha+1}\,.
\end{displaymath}
If the Laplace image is written as
\begin{displaymath}
\frac{s^{\alpha-1}}{s^\alpha+1}=\frac{1}{s}\,\frac{1}{(1+1/s^\alpha)}\,,
\end{displaymath}
the expression
\begin{displaymath}
\bar e_\alpha(s)+\frac{1}{s^\alpha}\,\bar e_\alpha(s)=\frac{1}{s}
\end{displaymath}
is treated by a (fractional) integral equation after the Laplace
transform. Since $e_\alpha(0)=1$, from the Laplace inversion we
find
\begin{equation}
e_\alpha(t)=1-J^\alpha e_\alpha(t)\,.\label{eq2}
\end{equation}
According to \cite{5}, the definition of fractional derivative is
\begin{displaymath}
\tilde D^\alpha u(t):=J^{m-\,\alpha}D^m u(t),
\end{displaymath}
taking the positive integer number $m$ with $m-1<\alpha\leq m$. In
fact, $D^m$ is the $m$-derivative under the integral
$J^{m-\,\alpha}$. Then the function $e_\alpha(t)$ satisfies also
the equivalent equation in the differential form
\begin{equation}
\tilde D^\alpha e_\alpha(t)+e_\alpha(t)=0\,.\label{eq3}
\end{equation}
It should be pointed out, though this equation was known earlier,
it was not derived but postulated. Using the approach, now we
intend to derive an equation describing $i_\alpha(t)$.

In order to obtain the equation for $i_\alpha(t)$, we need to
consider its Laplace image
\begin{displaymath}
i_\alpha(t)\quad\div\quad\frac{1}{s}\,\frac{1}{(s^{\alpha/2}+s^{-\,\alpha/2})}
\,.
\end{displaymath}
The initial condition is $i_\alpha(0)=0$. Thus, the corresponding
equation becomes
\begin{equation}
\tilde D^{\alpha/2}i_\alpha(t)+
J^{\alpha/2}i_\alpha(t)=1\,.\label{eq4}
\end{equation}
It is interesting to observe that following the argument of
\cite{1b}, the generalized momentum of the fractional oscillator
takes the form
\begin{displaymath}
p_\alpha=m\,\tilde D^{\alpha/2}
q_\alpha(t)=-m\,q_\alpha(0)\,\omega^2\,t^{\alpha/2}
E_{\alpha,1+\alpha/2}(-\omega^2t^\alpha)\,,
\end{displaymath}
where $q_\alpha$ is the displacement, $\omega$ the circular
frequency, $m$ the generalized mass. To put it in another way, the
momentum is expressed in terms of $-i_\alpha(t)$. Really,
expanding $e_\alpha(t)$ and $i_\alpha(t)$ in an infinite power
series about $t^\alpha$ and differentiating the functions by
$\tilde D^{\alpha/2}$ with respect to $t$ , we obtain
\begin{displaymath}
\tilde D^{\alpha/2}e_\alpha(t)=-i_\alpha(t)\,,\qquad \tilde
D^{\alpha/2}i_\alpha(t)=e_\alpha(t)\,,
\end{displaymath}
whereas
\begin{displaymath}
J^{\alpha/2}e_\alpha(t)=i_\alpha(t)\,,\qquad
J^{\alpha/2}i_\alpha(t)=1-e_\alpha(t)\,.
\end{displaymath}
Consequently, Eq.(\ref{eq4}) is reduced to
\begin{displaymath}
\tilde D^\alpha i_\alpha(t)+i_\alpha(t)=0\,.
\end{displaymath}
The form of this equation coincides with Eq.(\ref{eq3}). Thus, the
functions $e_\alpha(t)$ and $i_\alpha(t)$ represent two
independent solutions of the same equation.

Moreover, this permits one to derive the equation of fractional
oscillator by means of the Hamilton formalism with some
alterations. So the generalized Hamiltonian is written as
\begin{equation}
\mathcal{H}_\alpha=(p^2_\alpha+\omega^2q^2_\alpha)/2\,.\label{eq4a}
\end{equation}
It corresponds to the total energy of the fractional oscillator
\cite{1b}. In the Hamilton form the motion equation of fractional
oscillator becomes
\begin{eqnarray}
\tilde D^{\alpha/2}q_\alpha&=&\frac{d^{\alpha/2}q_\alpha}{d
t^{\alpha/2}}=\frac{\partial\mathcal{H}_\alpha}{\partial
p_\alpha}=p_\alpha\,,\label{eq5}\\ \tilde
D^{\alpha/2}p_\alpha&=&\frac{d^{\alpha/2}p_\alpha}{d
t^{\alpha/2}}=-\frac{\partial\mathcal{H}_\alpha}{\partial
q_\alpha}=-\omega^2q_\alpha\,.\label{eq6}
\end{eqnarray}
The modification of the conventional representation of the
Hamilton equations is conditioned on a random interaction of the
system of harmonic oscillators with environment. Each harmonic
oscillator is governed by its own internal clock. Although its
dynamics is described by the ordinary Hamiltonian equations,
$p(\tau)$ and $q(\tau)$ depend on the operational time. The
passage from the operational time to the physical time via the
averaging procedure accounts for the interaction of the system of
harmonic oscillators with environment. This results in this system
on the whole behaving as a fractional oscillator. The momentum of
the fractional oscillator and its displacement  take the integral
form
\begin{displaymath}
p_\alpha(t)=\int_0^\infty p^{S}(t,\tau)\,p(\tau)\,d\tau\,,\qquad
q_\alpha(t)=\int_0^\infty p^{S}(t,\tau)\,q(\tau)\,d\tau\,.
\end{displaymath}
Therefore, the fractional oscillator is an ensemble average of
harmonic oscillations (see details in \cite{2}). Although the
Hamiltonian (\ref{eq4a}) is not an explicit function of time, for
non-integer values $\alpha$ the dynamic system is nonconservative
because of the fractional derivative of momentum.

As is well known \cite{6}, the Laplace images are also useful for
finding an asymptotic behavior of transformed functions. With this
in mind we expand the term $1/(s^\alpha+1)$ in a series with
respect to $s$. This technique works finely for the solutions of
linear fractional differential equations \cite{1,7,8}. Taking the
leading terms, we obtain the asymptotic representations as
$t\to\infty$, namely
\begin{displaymath}
e_\alpha(t)\sim\frac{t^{-\,\alpha}}{\Gamma(1-\alpha)}\,,\qquad\qquad\qquad
i_\alpha(t)\sim\frac{t^{-\,\alpha/2}}{\Gamma(1-\alpha/2)}\,.
\end{displaymath}
For $1<\alpha<2$ the denominator $\Gamma(1-\alpha)$ is less than
zero, whereas $\Gamma(1-\alpha/2)>0$. It is easy to show that the
asymptotic algebraic decay of $e_\alpha(t)$ and $i_\alpha(t)$ is
completely determined by the parts $f_\alpha(t)$ and
$h_\alpha(t)$, respectively (see Section~\ref{par3}). In the
classical case $\alpha=2$ the contributions $f_\alpha(t)$ and
$h_\alpha(t)$ are equal to zero exactly, because there is only a
pole, and the cut on the negative real axis is no longer present.

\section{Zeros of fractional oscillations}\label{par5}

The fractional oscillations occupy a special place in the
oscillatory theory and what is why. On the one hand, they have a
decay like a relaxation. On the other hand they show themselves as
oscillations. Asymptotically the fractional oscillations become
vanishingly small with $t\to\infty$. However, the feature has a
deep cornerstone. Their representation in the form of two
contributions is especially important. The decomposition manifests
a competition between two different dependencies. The harmonic
oscillation exponentially decays on the background of a slow
algebraic relaxation. Although at the beginning the oscillations
prevail, the algebraic decay will survive them. As a result, the
fractional oscillations have a finite number of damped
oscillations as well as zeros.

Consider the problem in greater detail. Probably, Wiman was the
first who has described the position of the Mittag-Leffler
function zeros in the complex plane \cite{9}. A finite number of
the zeros has been established. Later the analysis was specified
in \cite{1}. The function $e_\alpha(t)$ demonstrates an odd number
of zeros. The smallest zero lies in the interval
$0<t<\pi/[\sin(\pi/\alpha)]$. The number of zeros strictly depends
on the index $\alpha$. Increasing $\alpha$ more and more, we get
more the number of zeros in the function $e_\alpha(t)$. For
$\alpha=1$ the only zero is located in infinity. When $\alpha$
tends from 1 to 2, the number of zeros increases so that for
$\alpha=2$ their number becomes equal to infinity.

We provide a similar analysis to the function $i_\alpha(t)$. It is
not evident that the function has a finite number of zeros for
$1<\alpha<2$\,. At once it should be noticed that this function
always is zero for $t=0$ under any value of $1\leq\alpha\leq 2$\,.
When $t$ is enough large, the zeros of $i_\alpha(t)$ are expected
to be found approximately from the equation
\begin{equation}
\frac{2}{\alpha}\,e^{\,t\cos(\pi/\alpha)}\approx\frac{t^{-\,\alpha/2}}
{\Gamma(1-\alpha/2)}\,,\label{eq7}
\end{equation}
neglecting the oscillation factor in the contribution
$q_\alpha(t)$. Putting now $\alpha=1+\varepsilon$, the first-order
approximation gives
\begin{displaymath}
\cos(\pi/\alpha)=\cos[\pi/(1+\varepsilon)]\sim\cos[\pi(1-\varepsilon)]\sim
-1\,.
\end{displaymath}
Thus, the asymptotic position $T$ of the largest zero is defined
by the relation
\begin{equation}
e^{-T}\sim\frac{(1+\varepsilon)}{2\Gamma(1/2+\varepsilon/2)}
T^{-1/2-\varepsilon/2}\,,\label{eq8}
\end{equation}
which shows that $T$ tends to infinity as $\varepsilon\to 0$.

The function $i_2(t)$ has infinitely many zeros. Now let the index
$\alpha$ be $2-\delta$. In the limit of the first-order
approximation we can write
\begin{displaymath}
\cos(\pi/\alpha)=\cos[\pi/(2-\delta)]\sim\cos[(\pi/2)(1+\delta/2)]=
\sin(\pi\delta/4)\sim\pi\delta/4
\end{displaymath}
and
\begin{displaymath}
\Gamma(1-\alpha/2)=\Gamma(\delta/2)\sim 2/\delta\,.
\end{displaymath}
In this case the asymptotic estimation for the largest zero $T$
reduces to the equation
\begin{displaymath}
e^{-\pi\delta T/4}\sim\frac{\delta(2-\delta)}{4}
T^{-1+\delta/2}\,,
\end{displaymath}
from which one gets
\begin{equation}
\pi\delta T/4\sim\ln(2T/\delta)\,.\label{eq9}
\end{equation}
Since $\delta\to 0$, the value $T$ tends to infinity faster than
$1/\delta$. Moreover, the terms $\pi\delta T/4$ and
$\ln(2T/\delta)$ are of the same order. Taking either
$T\sim(a/\delta)\ln(2/\delta)$ or $\delta\sim b\ln(T)/T$, where
$a$ and $b$ are positive constants to be determined, we arrive at
$a=b=8/\pi$. The equivalent asymptotic expressions are
\begin{eqnarray}
T&\sim&\frac{8}{\pi\delta}\,\ln\left(\frac{2}{\delta}\right)\,, \label{eq10}\\
\delta&\sim&\frac{8}{\pi}\,\frac{\ln T}{T}\,.\label{eq11}
\end{eqnarray}
In the limit $\alpha\to 2$ the period of $q_\alpha(t)$ becomes
$2\pi$. Consequently the number of zeros of $i_\alpha(t)$ tends to
$N\sim T/\pi\to\infty$ as $\delta\to 0$.

\section{Conclusions}

We have considered the fractional oscillations allied with
ordinary harmonic ones. The main feature of the fractional
oscillations is that they have a finite number of zeros. This
follows from the competition of two terms. One of them has an
asymptotic behavior with an algebraic decay, and the other term
contains a conventional harmonic oscillation vanishing in time
because of an exponential decay. The second term decreases faster
than this happens for the term with an algebraic decay. This
important peculiarity of fractional oscillations is just reflected
on the title of the paper. The analysis supports a key role of
Mittag-Leffler functions. The fractional oscillations satisfy a
linear fractional differential equation. We have derived it, using
the method of the Laplace transform. Our analytical solutions are
confirmed by a numerical treatment.

\section*{Acknowledgements}
The author wants to thanks Prof. Narahari Achar and Prof. George
Zaslavsky for useful discussions on the subject.



\begin{thebibliography}{2004}
\bibitem{1}
R.~Gorenflo, F.~Mainardi, Fractional oscillations and
Mittag-Leffler functions, Proceedings of RAAM '96, Kuwait
University (1996) 193.
\bibitem{1a}
F.~Mainardi, Chaos, Soliton \& Fractals 7 (1996) 1461.
\bibitem{1b}
B.~N.~Narahari Achar, J.~W.~Hanneken, T.~Enck, T.~Clarke, Physica
A297 (2001) 361.
\bibitem{1c}
I.~M.~Sokolov, Phys.Rev. E63 (2001) 056111.
\bibitem{2}
A.~A.~Stanislavsky, Phys.Rev. E70 (2004) 051103.
\bibitem{2a}
I.~S.~Gradshtein, I.~M.~Ryzhik, Tables of Integrals, Series, and
Products, Acad. Press, New York, 1980.
\bibitem{2b}
M.~Abramowitz, I.~A.~Stegun, Handbook of Mathematical Functions,
Dover, New York, 1972.
\bibitem{2c}
K.~S.~Miller, S.~G.~Samko, Integr. Transf. and Spec. Func. 12(4)
(2001) 389.
\bibitem{3}
M.~M.~Meerschaert, H.-P.~ Scheffler, J. Appl. Probab. 41 (2004)
623.
\bibitem{4}
A.~A.~Stanislavsky, Theor. and Math. Phys. 138 (2004) 418.
\bibitem{5}
Yu.~Rabotnov, Creep problems in structural members, North-Holland,
Amsterdam, 1969, p.~129. Originally published in Russian as: {\it
Polzuchest' Elementov Konstruktsii}, Nauka, Moscow, 1966.
\bibitem{6}
G.~Doetsch, Introduction to the Theory and Application of the
Laplace Transformations, Springer-Verlag, Berlin, 1974.
\bibitem{7}
I.~Podlubny, Solutions of linear fractional differential
equations, In P.~Rusev, I.~Dimovski and V.~Kiryakova (eds.), {\it
Transform Methods and Special Functions, Sofia 1994}, Science
Culture Technology, Singapore, 1995, pp. 227-237.
\bibitem{8}
K.~S.~Miller, B.~Ross, An Introduction to the Fractional Calculus
and Fractional Differential Equations, Wiley, New York, 1993.
\bibitem{9}
A.~Wiman, Acta Math. 29 (1905) 217.
\end{thebibliography}
\end{document}